# Detailed Kinetic Study of the Ring Opening of Cycloalkanes by CBS-QB3 Calculations

B. Sirjean,[1] P.A. Glaude,[1] M.F. Ruiz-Lopez,[2] R. Fournet[1,*]

1) Département de Chimie Physique des Réactions, UMR 7630 CNRS-INPL, ENSIC, 1, rue Grandville, BP 20451, 54001 Nancy Cedex, France

2) Equipe de Chimie et Biochimie Théoriques, UMR CNRS-UHP No. 7565, Université Henri Poincaré – Nancy I, Boulevard des Aiguillettes, BP 239, 54506 Vandoeuvre-lès-Nancy, France

e-mail: Rene.Fournet@ensic.inpl-nancy.fr

ABSTRACT

This work reports a theoretical study of the gas phase unimolecular decomposition of cyclobutane, cyclopentane and cyclohexane by means of quantum chemical calculations. A biradical mechanism has been envisaged for each cycloalkane, and the main routes for the decomposition of the biradicals formed have been investigated at the CBS-QB3 level of theory. Thermochemical data ($\Delta H°_f$, $S°$, $C°_p$) for all the involved species have been obtained by means of isodesmic reactions. The contribution of hindered rotors has also been included. Activation barriers of each reaction have been analyzed to assess the energetically most favorable pathways for the decomposition of biradicals. Rate constants have been derived for all elementary reactions using transition state theory at 1 atm and temperatures ranging from 600 to 2000 K. Global rate constant for the decomposition of the cyclic alkanes in molecular products have been calculated. Comparison between calculated and experimental results allowed to validate the theoretical approach. An important result is that the rotational barriers between the conformers, which are usually neglected, are of importance in decomposition rate of the largest biradicals. Ring strain energies (RSE) in transition states for ring opening have been estimated and show that the main part of RSE contained in the cyclic reactants is removed upon the activation process.

**1. Introduction**

Many important gas-phase or heterogeneous industrial processes such as combustion, partial oxidation, cracking or pyrolysis exhibit a complex chemical scheme involving hundreds of species and several thousands of reactions. In these thermal processes, cyclic hydrocarbons, particularly cycloalkanes, represent an important class of compounds. These molecules are produced during the gas-phase processes though they can also be present in the reactants in large amounts; for example, a commercial jet fuel contains about 26% of naphtenes and condensed naphtenes, while in a commercial diesel fuel, this percentage reaches 40 % [1]. Modeling their reactivity currently represents an important challenge in the formulation of new fuels, less polluting and usable with new types of combustion in engines like "Homogeneous Charge Compression Ignition" (HCCI) [2]. During combustion or pyrolysis processes, cycloalkanes can lead to the formation of a) toxic compounds or soot precursors such as benzene (by successive dehydrogenations) and b) linear unsaturated species such as buta-1,3-diene or acroleïn (by the opening of the ring). Several experimental and modeling studies have been carried out on the oxidation of cyloalkanes in gas phase [3-9]. However, the modeling of the combustion of cycloalkanes remains difficult due to the lack of both, kinetic data for elementary reactions and thermodynamic data ($\Delta H°_f$, $S°$, $C°_p$) for the relevant species.

A number of experimental and theoretical works have been reported on the decomposition of small cycloalkanes, such as cyclopropane [10-16]. Ring opening in this molecule is a well-known process and the rate of dissociation leading to propene formation has been extensively measured and calculated. Not less than 37 estimations of the rate constant are available on the NIST chemical kinetics database [17]. Thermal decomposition of cyclobutane has been experimentally studied too and rate constants for the ring opening leading to the formation of two ethylene molecules have been measured [15-16, 18-19]. Theoretical studies on cyclobutane have mainly focused on the reverse reaction, namely, the cycloaddition of two ethylene molecules [20-23], since it represents a prototype reaction for the Woodward-Hoffmann rules and illustrates the usefulness of orbital symmetry considerations. Moreover, Pedersen et al. [24] have showed the validity of the biradical hypothesis by direct femtosecond studies of the transition-state structures. Theses studies have highlighted the fact that cycloaddition of two ethylene molecules can proceed through two different routes: one involves a tetramethylene biradical intermediate, while the other implies a concerted reaction that directly leads to cyclobutane formation. However, the latter reaction has been shown to have a high activation energy due to steric effects (see for instance ref [20] and therein) and to be much less favorable than the biradical process.

Even though the study of ring opening of cyclopropane and cyclobutane are interesting from a theoretically point of view, larger cycloalkanes such as cyclopentane or cyclohexane are mainly involved in usual fuel. In spite of that, the unimolecular initiation of these molecules has been little investigated. Ring opening of cyclopentane and cylohexane has been experimentally studied by Tsang [25-26] who has reported the main routes of decomposition of these molecules. The mechanism and initial rates of decomposition were determined from single-pulse shock-tube experiments. For cyclohexane, a reaction pathway leading to the formation of 1-hexene has been considered only whereas for cyclopentane, the processes leading to either 1-pentene or to cyclopropane + ethylene have both been investigated, in accordance with the products experimentally detected. These results have been confirmed by further experimental studies by Kalra et al. [27] and Brown et al. [28]. In addition, Tsang [25] has shown that the experimentally obtained global rate parameters are consistent with a biradical mechanism for ring opening (**Scheme 1**):

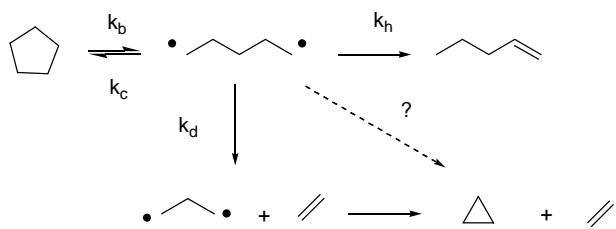

**Scheme 1.** Biradical mechanism for cyclopentane ring opening [25]

Tsang connected the global rate parameters for the decomposition of cyclopentane in 1-pentene ($k_1$) or in cyclopropane+ethylene ($k_2$) to the rate parameters of the elementary reactions shown in Scheme 1. The equilibrium constant $K_{eq}=k_b/k_c$ has been estimated by the group additivity method proposed by Benson [29] and using some analogies with reactions of linear alkanes. From these assumptions, the rate constants of the elementary reactions have been estimated. The same procedure has been applied for the rate of decomposition of cyclohexane in 1-hexene. Even if the rate parameters obtained for the elementary reactions are consistent with the formation of a biradical, no transition state has been defined for validating the suggested mechanism. Moreover, the route leading to ethylene and trimethylene through C-C bond cleavage is rather ambiguous since other pathways are considered by Tsang, as for instance the direct decomposition of the *n*-pentyl biradical into cyclopropane and ethylene (Scheme 1). All of these studies showed that biradicals are central to the understanding of reaction mechanisms as well as to the predictability of reaction products and rates in the ring opening of cycloalkanes.

Our aim in the present work is to analyze the ring opening of cyclobutane, cyclopentane and cyclohexane by means of high level quantum chemical calculations in order to obtain accurate rate constants for elementary reactions. Comparison of the computed rate constants with available experimental data allows us to validate the theoretical approach. We explore several plausible pathways that could be involved in the decomposition of the biradicals and we discuss the evolution of the ring strain energy (RSE) in going from the reactants to the transition state for ring opening.

**2. Computational method**

Quantum chemical computations have been performed on an IBM SP4 computer with the Gaussian03 (G03) software package [30]. The high-level composite method CBS-QB3 [31] has been used. Analysis of vibrational frequencies confirmed that all transition structures (TS) have one imaginary frequency. Intrinsic Reaction Coordinate (IRC) calculations have been systematically performed at the B3LYP/6-31G(d) [32-33] level to ensure that the computed TSs connect the desired reactants and products. Only singlet biradicals states have been considered and their study at the composite CBS-QB3 level requires two specific comments. First, at this level, geometry optimization of the systems is performed by density functional theory (DFT) using an unrestricted B3LYP method and a CBSB7 basis set. It is worth noting that the use of one determinantal wavefunction to describe open shell singlet biradicals can be questionable. However, previous studies have shown that the geometries obtained in this way compare well with those obtained at more refined computational levels [34-38]. Second, in CBS-QB3 calculations, a correction for spin-contamination in open-shell species is added to the total energy. It includes a term of the form $\Delta E = -0.00954 \left( S^2 - S^2_{th} \right)$ where $S^2$ denotes the expected value of the $\hat{S}^2$ operator and $S^2_{th}$ the corresponding theoretical value (e.g. 0 for a singlet state). This correction was derived for systems displaying a small spin-contamination. However, because of strong singlet-triplet mixing in the unrestricted wavefunction for biradicals, the $S^2$ values are close to 1 and this leads to a systematic error in the CBS-QB3 energy correction of about 6 kcal.mol$^{-1}$. Several papers have pointed out this limitation [39-41] and some authors have proposed to remove the spin-contamination correction in this case. In the present work, we have preferred to use an empirical parameter specifically designed to handle spin-contamination in singlet biradicals so that singlet-triplet gaps for hydrocarbons biradicals are correctly described at the CBS-QB3 level. All energy values for singlet biradical species are therefore corrected by an expression of the form:

$$\Delta E_{spin} = -0.031\ (S^2 - S^2_{th}) \qquad (1)$$

with $S^2_{th} = 1$.

## 3. Thermochemical data

Thermochemical data ($\Delta H_f°$, $S°$, $C_p°$) for all the species involved in this study have been derived from energy and frequency calculations and are collected in **Table 1**. In the CBS-QB3 method, harmonic frequencies, at the B3LYP/CBSB7 level of theory, are scaled by a factor 0.99. Explicit treatment of the internals rotors has been performed with the *hinderedRotor* option of Gaussian03 in accordance with the work of Ayala and Schlegel [42]. A systematic analysis of the results obtained was made in order to ensure that internals rotors were correctly treated. Thus, for biradicals a practical correction was introduced in order to take into account the symmetry number of 2 for each $CH_2(\bullet)$ terminal group. This symmetry is not automatically recognized by Gaussian in the case of a radical group. Moreover, it must be stressed that in transition states, the constrained torsions of the cyclic structure have been treated as harmonic oscillators and the free alkyl groups as hindered rotations.

Enthalpies of formation ($\Delta H_f°$) of species involved in this study have been calculated using isodesmic reactions [43] excepted for cyclanes and 1-alkene for which more accurate experimental enthalpies of formation [44] can be found in the literature. Thanks to the conservation of the total number and types of bonds, very good results can be obtained due to the cancellation of errors on the two sides of the reaction. Several isodesmic reactions have been considered for the calculation of $\Delta H_f°$ in order to obtain an average value. However, results appear to be strongly dependent on the accuracy of the experimental data used for species involved in isodesmic reactions, especially for biradicals.

**Table 1**. Ideal gas phase thermodynamics properties obtained by CBS-QB3 calculation. $\Delta H^\circ_{f,298K}$ is expressed in kcal.mol$^{-1}$ and $S^\circ_{298K}$ and $C^\circ_p(T)$ are given in cal.mol$^{-1}$.K$^{-1}$.

| Species | $\Delta H^\circ_{f,298K}$ | $S^\circ_{298K}$ | $C^\circ_p(T)$ | | | | | | | P.G. |
|---|---|---|---|---|---|---|---|---|---|---|
| | | | 300 K | 400 K | 500 K | 600 K | 800 K | 1000 K | 1500K | |
|  | 12.74 | 56.78 | 13.28 | 18.03 | 22.25 | 25.74 | 31.04 | 34.92 | 40.98 | D$_{3h}$ |
|  | 6.79 | 63.19 | 16.97 | 23.47 | 29.37 | 34.33 | 41.94 | 47.48 | 55.94 | D$_{2d}$ |
|  | -18.27 | 70.21 | 20.99 | 29.17 | 36.67 | 43.01 | 52.81 | 59.92 | 70.74 | D$_{5h}$ |
|  | -29.43 | 71.55 | 25.45 | 35.27 | 44.28 | 51.96 | 63.89 | 72.55 | 85.66 | D$_{3d}$ |
|  | -22.99 | 74.09 | 25.61 | 35.33 | 44.30 | 51.95 | 63.84 | 72.49 | 85.60 | D$_2$ |
| 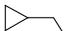 | 0.82 | 78.27 | 24.52 | 32.14 | 38.82 | 44.39 | 53.03 | 59.41 | 69.39 | C$_1$ |
| 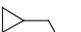 | -4.40 | 84.87 | 29.11 | 39.91 | 47.71 | 54.90 | 65.51 | 73.16 | 84.94 | C$_1$ |
| 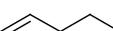 | -5.09 | 79.85 | 25.99 | 32.74 | 38.83 | 44.05 | 52.37 | 56.64 | 68.53 | C$_1$ |
| 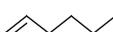 | -9.90 | 91.75 | 31.26 | 39.41 | 46.75 | 53.05 | 63.11 | 70.69 | 82.62 | C$_1$ |
| 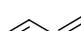 | 26.31 | 66.32 | 18.06 | 22.98 | 27.35 | 31.02 | 36.58 | 40.52 | 46.70 | C$_{2h}$ |
| 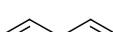 | 25.41 | 78.56 | 23.57 | 29.58 | 34.92 | 39.42 | 46.46 | 51.70 | 59.91 | C$_2$ |
| 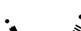 | 67.43 | 79.01 | 22.95 | 27.80 | 32.16 | 35.90 | 41.92 | 46.54 | 53.94 | C$_1$ |
| 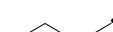 | 68.13 | 77.33 | 23.59 | 28.36 | 32.50 | 36.10 | 41.93 | 46.46 | 53.82 | C$_{2h}$ |
| 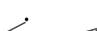 | 62.84 | 87.11 | 27.87 | 34.29 | 40.04 | 44.93 | 52.73 | 58.64 | 68.05 | C$_1$ |
| 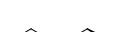 | 62.46 | 87.12 | 28.37 | 34.50 | 40.11 | 44.95 | 52.71 | 58.62 | 68.03 | C$_1$ |
| 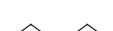 | 62.79 | 89.54 | 28.15 | 34.26 | 39.89 | 44.75 | 52.52 | 58.44 | 67.90 | C$_1$ |
| 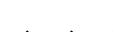 | 58.50 | 98.74 | 31.50 | 39.48 | 46.76 | 52.98 | 62.83 | 70.22 | 81.87 | C$_1$ |
| 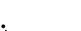 | 59.41 | 94.29 | 33.61 | 41.17 | 48.03 | 53.96 | 63.47 | 70.70 | 82.15 | C$_1$ |
| 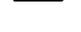 | 59.08 | 92.96 | 34.02 | 41.97 | 49.00 | 54.97 | 64.47 | 71.58 | 82.74 | C$_1$ |

Hence, for these systems, $\Delta H_f°$ has been obtained from the prototype reaction:

$$\bullet R_1\text{-}\bullet R_2 + 2 R_3H \rightarrow HR_1\text{-}R_2H + 2 \bullet R_3, \tag{2}$$

where $\bullet R_1\text{-}\bullet R_2$ represents the biradical and $\bullet R_3 = \bullet H$, $\bullet CH_3$, $\bullet C_2H_5$ or n-$\bullet C_3H_7$.

The computed entropy of cyclopentane has been corrected in order to take into account the experimental symmetry of the molecule ($D_{5h}$). Indeed, CBS-QB3 calculations lead to a non planar geometry with $C_1$ symmetry. In addition, a low frequency of 22 cm$^{-1}$ can be associated with a puckering motion of the ring. According to Benson [29], this pseudo-rotation is so fast that cyclopentane can be treated as dynamically flat (with a symmetry number $\sigma=10$). Thus, we corrected the entropy of cyclopentane by R ln10 (equation 3):

$$S_{c\text{-}C5H10} = S_{c\text{-}C5H10}(\text{CBS-QB3}) - R \ln 10 \tag{3}$$

The computed value, 70.0 cal.mol$^{-1}$.K$^{-1}$, is in good agreement with the experimental value [29].

To our knowledge, no experimental enthalpy of formation for biradicals $\bullet C_4H_8\bullet$, $\bullet C_4H_{10}\bullet$ or $\bullet C_6H_{12}\bullet$ has been reported. However, we can compare our values with those obtained from an estimation based on bond dissociation energy (BDE) according to reaction (4):

$$C_nH_{2n+2} = \bullet C_nH_{2n}\bullet + 2\, H\bullet \tag{4}$$

where $C_nH_{2n+2}$ represents a linear free alkane and $\bullet C_nH_{2n}\bullet$ is the corresponding biradical. $\Delta H°_r$ for reaction 4 corresponds to twice the BDE for a C-H bond and the enthalpy of formation of $\bullet C_nH_{2n}\bullet$ can therefore be calculated from equation (5):

$$\Delta H°_f (\bullet C_nH_{2n}\bullet, 298K) = 2\,BDE + \Delta H°_f (C_nH_{2n+2}, 298K) - 2\, \Delta H°_f (H\bullet, 298K) \tag{5}$$

This calculation rests on the assumption that no interaction exists between the two radical centers.

**Table 2** compares estimated values using equation (5) and CBS-QB3 results for the most stable conformation of the biradicals $\bullet C_4H_8\bullet$, $\bullet C_4H_{10}\bullet$ or $\bullet C_6H_{12}\bullet$. The BDE value for a primary C−H bond in a linear alkane has been taken equal to 100.9 kcal.mol$^{-1}$, as proposed by Tsang [45] (it corresponds to C-H bond dissociation in propane to give the n-propyl radical). Experimental enthalpies of formation for the molecules used in equation (5) come from NIST webbook [44]. As shown in Table 2, a very good agreement is obtained between CBS-QB3 calculations using isodesmic reactions and values estimated by BDE and equation (5). This result corroborates the consistency of the electronic calculation scheme for the biradicals, in particular, the use of a broken symmetry method to optimize their geometry, as discussed in Section 2.

**Table 2**. Comparison of $\Delta H^{\circ}_f$ (in kcal.mol$^{-1}$, at 298 K) estimated from equation (5) and from theoretical calculations.

| Biradical | $\Delta H^{\circ}_f$ from equation (5) | This work |
|---|---|---|
| •C$_4$H$_8$• | 67.57 | 67.43 |
| •C$_5$H$_{10}$• | 62.52 | 62.46 |
| •C$_6$H$_{12}$• | 57.64 | 58.50 |

**4. Kinetic calculations**

The rate constants involved in the mechanisms were calculated using TST [47]:

$$k_{uni} = rpd\ \kappa(T) \frac{k_b T}{h} \exp\left(\frac{\Delta S^{\neq}}{R}\right) \exp\left(-\frac{\Delta H^{\neq}}{RT}\right) \qquad (6)$$

where $\Delta S^{\neq}$ and $\Delta H^{\neq}$ are, respectively, the entropy and enthalpy of activation and *rpd* is the reaction path degeneracy. For reactions involving H-transfer, a transmission coefficient, namely $\kappa(T)$, has been calculated in order to take into account tunneling effect. We used an approximation to κ, provided by Skodje and Truhlar [47]:

$$\begin{aligned}
&\textbf{for } \beta \leq \alpha \\
&\kappa(T) = \frac{\beta\pi/\alpha}{\sin(\beta\pi/\alpha)} - \frac{\beta}{\alpha - \beta} e^{\left[(\beta-\alpha)(\Delta V^{\neq} - V)\right]} \\
&\textbf{for } \alpha \leq \beta \\
&\kappa(T) = \frac{\beta}{\beta - \alpha}\left[e^{\left[(\beta-\alpha)(\Delta V^{\neq} - V)\right]} - 1\right]
\end{aligned} \qquad (7)$$

where $\alpha = \dfrac{2\pi}{h\,\mathrm{Im}(v^{\neq})}, \beta = \dfrac{1}{k_B T}$

In equation (7), $v^{\neq}$ is the imaginary frequency associated with the reaction coordinate, $\Delta V^{\neq}$ is the zero-point-including potential energy difference between the TS structure and the reactants, and V is 0 for an exoergic reaction and the zero-point-including potential energy difference between reactants and products for an endoergic reaction.

The calculation of $\Delta H^{\neq}$ was performed by taking into account electronic energies of reactants and TSs but also the enthalpy of reaction (**Scheme 2**).

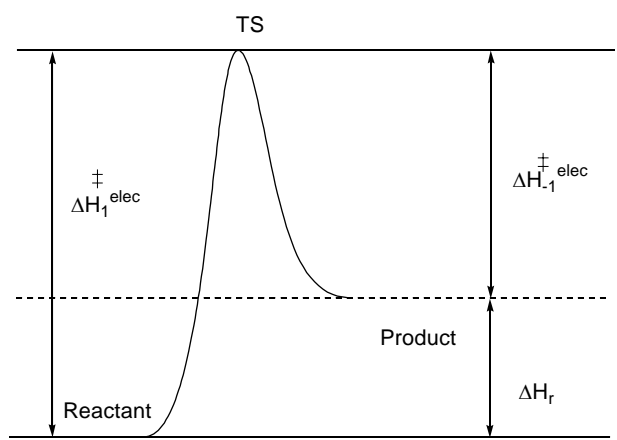

**Scheme 2.** Calculation of enthalpy of reaction: $\Delta H^{\neq}$

Thus, $\Delta H^{\neq}$ (reactant → product) is calculated from the following relation:

$$\Delta H^{\neq}(R \to P) = \left(\Delta H_1^{\neq}(elec) + \Delta H_{-1}^{\neq}(elec) + \Delta H_r\right)/2 \qquad (8)$$

and $\Delta H^{\neq}$ (product → reactant) by:

$$\Delta H^{\neq}(P \to R) = \left(\Delta H_1^{\neq}(elec) + \Delta H_{-1}^{\neq}(elec) - \Delta H_r\right)/2 \qquad (9)$$

where $\Delta H_1^{\neq}(elec)$ and $\Delta H_{-1}^{\neq}(elec)$ are, respectively, the enthalpy of activation for the direct and reverse reactions, calculated from electronic energies, ZPVE and thermal correction energy. $\Delta H_r$ corresponds to the enthalpy of reaction estimated using isodesmic reactions [48]. The use of equations (8) and (9) ensures the consistency between kinetic and thermodynamic data and improves the activation enthalpy results since the values obtained for $\Delta H_r$ are more accurate than direct electronic calculations.

The kinetic data are obtained with a fitting of equation 6 in the temperature range 600-2000K, with the following modified Arrhenius form:

$$k = A\, T^b \exp(-E/RT) \qquad (10)$$

## 5. Ring opening mechanisms

The ring opening mechanisms for cyclobutane, cyclopentane and cyclohexane are now discussed. In all schemes presented below, electronic free enthalpies are reported in kcal.mol$^{-1}$ and are relative to the reference cycloalkane at P=1 atm. The value in bold corresponds to free enthalpy at T=298 K while the value in italics corresponds to T=1000 K.

### 5.1 Cyclobutane

**Scheme 3** presents the global mechanism obtained for the ring opening of cyclobutane. In this scheme, we only considered conformers (molecules, biradicals, TSs) of lowest free enthalpies. Especially, we do not account for conformations of the biradical (2) (*gauche* and *trans*) and we only considered the *gauche* conformer since its free enthalpy is lowest. This point will be discussed in more detail below.

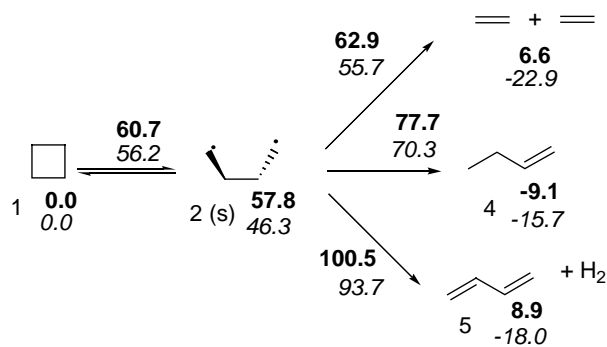

*(s) singlet state*

*$\Delta G°(T)$ in kcal.mol$^{-1}$ (bold : T= 298K, italic : T=1000K)*

**Scheme 3.** Mechanism obtained for the ring opening of cyclobutane and for conformers of lowest energies.

The ring opening of cyclobutane involves a low free enthalpy of activation; $\Delta G^{\neq}$ = 60.7 kcal.mol$^{-1}$ at 298 K which corresponds to an activation enthalpy ($\Delta H^{\neq}$) of 62.7 kcal.mol$^{-1}$. This value is lower than that involved in the dissociation of a C-C bond in the corresponding linear alkane ($\Delta H^{\neq} \approx$ 86.3 kcal.mol$^{-1}$) [49]. This result will be discussed below but we can anticipate that it is mainly due to the high ring strain energy involved in the cyclobutane molecule, that is partially removed in the transition state. Following ring opening, three ways of decomposition for tetramethylene have been investigated. The route leading to the formation of two ethylene molecules corresponds to the most favorable one, as already mentioned in the literature [15-23]. The process leading to 1-butene represents another possible pathway though this elementary reaction involves a high stressed cyclic transition state for H-transfer (four-member ring) and displays a high activation energy ($\Delta H^{\neq}$ =17.6 kcal.mol$^{-1}$ vs 2.8 kcal.mol$^{-1}$ for the β-scission leading to $C_2H_4$, at 298 K). This pathway, however, might provide a non–negligible contribution to the total process. Indeed, $C_4H_8$ has been identified as a minor product in experimental studies of cyclobutane decomposition [18]. The third route consists of the abstraction of two hydrogen atoms to form buta-1,3-diene and $H_2$. The reaction step has a high activation barrier ($\Delta H^{\neq}$ = 40 kcal.mol$^{-1}$ at T=298K) and should not compete with the previous mechanisms.

A concerted reaction leading from cyclobutane to direct formation of two ethylene molecules has been envisaged but no consistent TS could be obtained. As mentioned previously, several studies [20-23] have been performed on the reverse reaction, i.e. the cyclic addition of two ethylene molecules. The authors found that a two-step biradical-type reaction is expected to be favored (by about 24 kcal.mol$^{-1}$) over concerted pathways. Sakai [20] estimated activation energies of 77.6 kcal.mol$^{-1}$ and 58.8 kcal.mol$^{-1}$ for the concerted and stepwise processes, respectively, at the MP2//CAS/6-311+G(d,p) level. However, it is important to underline that this author reported a second-order saddle point structure (two negative eigenvalues) for the concerted mechanism, which therefore does not correspond to a transition state.

Table 3 gives the kinetics parameters of the modified Arrhenius form (equation 10) for all the elementary processes involved in Scheme 3. Unimolecular initiation of cycloalkanes is unimportant for low temperatures in thermal processes and we only give kinetics parameters for T > 600K in all the tables.

Table 3. Rate parameters for the unimolecular initiation of cyclobutane at P=1 atm, 600 ≤ T (K) ≤ 2000 K and related to scheme 3.

|  | $k_{1-2}$ | $k_{2-1}$ | $k_{2-C2H4}$ | $k_{2-4}$ | $k_{2-5}$ |
|---|---|---|---|---|---|
| log A (s$^{-1}$) | 18.53 | 12.21 | 7.32 | 5.57 | 2.23 |
| n | -0.797 | -0.305 | 1.443 | 2.171 | 2.995 |
| $E_a$ (kcal.mol$^{-1}$) | 64.85 | 1.98 | 3.03 | 16.44 | 37.61 |

To the best of our knowledge, no previous quantum calculations have been performed to estimate rate parameters for the elementary reactions involved in the thermal decomposition of cyclobutane. In 1971, Beadle et al. [19] experimentally studied the pyrolysis of cyclobutane and reported rate parameters for the ring opening of c-$C_4H_8$ ($k_{1-2}$ and $k_{2-1}$) and for the decomposition of the biradical in $C_2H_4$ ($k_{2-C2H4}$). Their estimations are based on tabulated thermochemistry and additivity methods. The A factor and activation energy proposed by Beadle et al. are as follows: 3.63 10$^{15}$ s$^{-1}$ and 63.34 kcal.mol$^{-1}$ for $k_{1-2}$, 2 10$^{12}$ s$^{-1}$ and 6.6 kcal.mol$^{-1}$ for $k_{2-1}$, and 1.17 10$^{13}$ s$^{-1}$ and 8.25 kcal.mol$^{-1}$ for $k_{2-C2H4}$. In spite of considerable uncertainty in that study, as mentioned by the authors, their results are in good agreement with ours. Thus, at 800 K, the ratio between our value and that proposed by Beadle et al. is 1.7, 2.0 and 0.7 for $k_{1-2}$, $k_{2-1}$, and $k_{2-C2H4}$, respectively. Moreover, it is worth noting that our CBS-QB3 quantum calculations provide an accurate energy for the cycloaddition of two ethylene molecules (formation of biradical 2). The free enthalpy of activation obtained is equal to 56.3 kcal.mol$^{-1}$ at 298K, which corresponds to an activation energy $\Delta H^{\neq} = 47.7$ kcal.mol$^{-1}$. This last value can be compared with the activation energy of 56.8 kcal.mol$^{-1}$ reported by Sakai [20] at the MP2/CAS/6-311+G(d,p) level though this author obtained a second order saddle point.

As mentioned previously, *gauche/trans* interconversion of (2) has been neglected in Scheme 3. However, the β-scission reaction involves a low ΔG$^{\neq}$ (5.1 kcal.mol$^{-1}$ at 298K) that could be of the same order of magnitude than the rotational barrier around the central CC bond. Accordingly, it can be interesting to consider the two conformations of the biradical for this particular but important pathway. A detailed scheme is presented in **Scheme 4**.

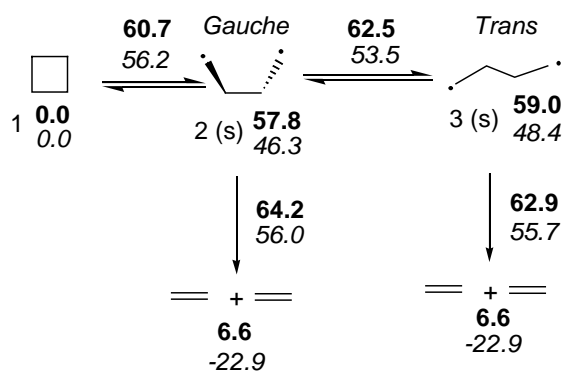

*(s) singlet state*
*ΔG°(T) in kcal.mol$^{-1}$ (bold : T= 298K, italic : T=1000K)*

**Scheme 4**. Detailed mechanism of C$_2$H$_4$ formation from the ring opening of cyclobutane and obtained by considering different conformers of C$_4$ biradicals.

Cyclobutane ring-opening leads to the *gauche* biradical (2). The latter can decompose in two C$_2$H$_4$ molecules or rotate to give the *trans* conformation (3), which in turn, can also decompose in two C$_2$H$_4$ molecules. The kinetic parameters for all processes involved in Scheme 4 are collected in **Table 4**.

**Table 4**. Rate parameters for the unimolecular initiation of cyclobutane at P=1 atm, $600 \leq T~(K) \leq 2000$ K and related to scheme 4.

|  | $k_{1-2}$ | $k_{2-1}$ | $k_{2-3}$ | $k_{3-2}$ | $k_{2\text{-}C2H4}$ | $k_{3\text{-}C2H4}$ |
|---|---|---|---|---|---|---|
| log A (s$^{-1}$) | 18.53 | 12.21 | 11.30 | 11.27 | 7.63 | 7.32 |
| n | -0.797 | -0.305 | 0.461 | 0.545 | 1.453 | 1.521 |
| $E_a$ (kcal/mol) | 64.85 | 1.98 | 4.29 | 3.26 | 4.79 | 2.07 |

Tables 3 and 4 give the kinetic parameters involved in the thermal decomposition of cyclobutane. In order to validate our results, we compare the global rate constant for the process:

$$\text{Cyclobutane} \rightarrow C_2H_4 + C_2H_4,$$

measured experimentally by Barnard et al. [15] and Lewis et al. [16], with that derived from our computations (Schemes 3 and 4). We consider the quasi-stationary state approximation (QSSA) for biradicals. For Scheme 3, QQSA applied to the *gauche* biradical leads to the simple expression:

$$k_g^{scheme3} = \frac{k_{1-2}\, k_{2-C_2H_4}}{k_{2-1} + k_{2-C_2H_4}}, \tag{11}$$

For scheme 4, QSSA leads to a more complex expression:

$$k_g^{scheme4} = \mathbf{k}_{3-C_2H_4} + k_{2-C_2H_4}\left(\frac{k_{3-2} + k_{3-C_2H_4}}{k_{2-3}}\right)\mathbf{A}, \tag{12}$$

with A= $\dfrac{k_{1-2}\, k_{2-3}}{k_{2-1}\, k_{3-2} + k_{2-1}\, k_{3-C_2H_4} + k_{2-3}\, k_{3-C_2H_4} + k_{2-C_2H_4}\, k_{3-2} + k_{2-C_2H_4}\, k_{3-C_2H_4}}$,

The fit of the global rate constants obtained from relations (11) and (12) leads to the rate parameters presented in **table 5.**

**Table 5**. Rate parameters for the global reaction cC$_4$H$_8$ → 2 C$_2$H$_4$, at P=1 atm, 600 ≤ T (K) ≤ 2000 K and related to Schemes 3 and 4.

|  | $k_g^{scheme3}$ | $k_g^{scheme4}$ |
|---|---|---|
| log A (s$^{-1}$) | 20.29 | 21.52 |
| n | -1.259 | -1.606 |
| E$_a$ (kcal/mol) | 67.69 | 68.16 |

**Figure 1** compares our results ($k_g^{scheme3}$ and $k_g^{scheme4}$) with the absolute value measured directly by Barnard et al. [15] and by Lewis et al. [16] for the thermal rate decomposition of cyclobutane in two ethylene molecules.

**Figure 1**. Comparison between calculated rate constant and experimental data for the global reaction cC$_4$H$_8$ → C$_2$H$_4$ + C$_2$H$_4$

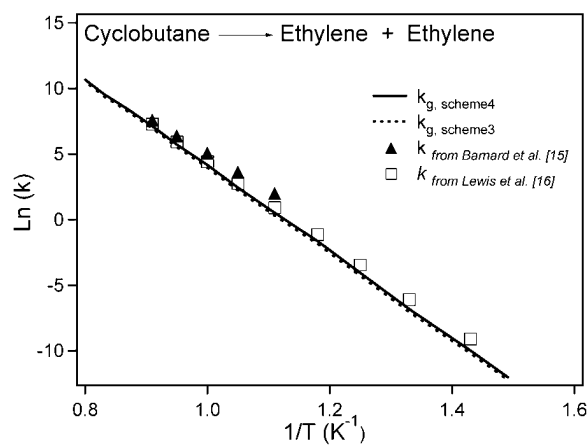

As shown in Figure 1, our results are close to the experimental values, validating the consistency of the theoretical approach. Computed rate constants are always slightly lower than those obtained by Lewis et al. [16] (maximum factor 2.4) or by Barnard et al. [15] (maximum factor 4.3). It is interesting to note that differences between $k_g^{scheme4}$ and $k_g^{scheme3}$ decrease with temperature, which is consistent with rotational hindrance (20% at 600K and 8% at 2000 K). Though these differences are weak, the rate constant obtained by explicit consideration of the two conformations of the biradical (*trans* and *gauche*) is closer to experimental results than the rate constant calculated from Scheme 3.

**5.2 Cyclopentane**

**Scheme 5** shows the global mechanism obtained for the ring opening of cyclopentane. As for cyclobutane, the global mechanism does not take into account the different conformations of the biradical •C$_5$H$_{10}$• and we only consider the conformation with the lowest free enthalpy. Due to weaker ring strain energy, the free enthalpy of activation of the ring opening of cyclopentane ($\Delta G^{\neq}$ = 80.5 kcal.mol$^{-1}$ at 298 K) is higher than that obtained in cyclobutane ($\Delta G^{\neq}$ = 60.7 kcal.mol$^{-1}$ at 298 K).

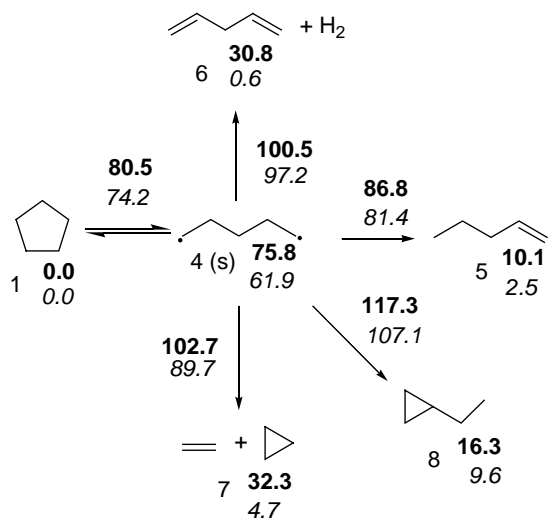

(s) singlet state

$\Delta G°(T)$ in kcal.mol$^{-1}$ (bold : T= 298K, italic : T=1000K)

**Scheme 5.** Mechanism obtained for the ring opening of cyclopentane and for conformers of lowest energies.

Four routes of decomposition have been investigated for the biradical 4, the most favorable one being the formation of 1-pentene by H-transfer. This last reaction exhibits an activation energy much lower than the equivalent process in cyclobutane ($\Delta H^{\neq}$ = 6 kcal.mol$^{-1}$ vs 17.6 kcal.mol$^{-1}$ for cC$_4$H$_8$ at 298 K).

This noticeable difference can be ascribed to different strain energy in the corresponding transition states. Conversely, the β-scission process leading to ethylene and cyclopropane is much more difficult than that leading to ethylene in cyclobutane. In this last case, the presence of two radical centers in (•$C_4H_8$•) in β position weakens the inner C-C σ-bond. Another interesting point concerns the dehydrogenation reaction of biradical 4 leading to penta-1,4-diene (6) and $H_2$. At low temperature (298K), this reaction is competitive with β-scission (reaction 4 →7) but becomes unimportant at high temperature (1000K) due to a low change of entropy between the biradical and TS. Finally, the reaction of biradical (4) to yield ethyl-cyclopropane (8) is unlikely to occur at either room or high temperature because it involves a high stressed cyclic transition state for H-abstraction and always displays quite high activation energy. For completeness, it must be noted that formation of a trimethylene biradical from 4 has been envisaged. However, the singlet state for this system could not be obtained, geometry optimizations leading systematically to the formation of cyclopropane (only the triplet state could be optimized). This may be due to the monodeterminantal character of the methodology used for the geometry optimizations but it has been shown that ring closure of trimethylene to form cyclopropane occurs very fast anyway [50]. **Table 6** gives the kinetics parameters of the modified Arrhenius forms (equation 7) for all the elementary processes involved in Scheme 5. As mentioned previously, just a few studies have been devoted to the kinetics of the thermal decomposition of cyclopentane [25, 27]. Tsang [25] measured by comparative-rate-pulse shock tube experiments, the ring opening of cyclopentane, and suggested rate expressions for $k_{1-4}$, $k_{4-1}$, $k_{4-5}$ and $k_{4-7}$, over the temperature range 1000 K – 1200 K. The estimations were based on experimental results obtained for the thermal decomposition of cyclopentane and on the thermochemical considerations methods proposed by Benson [29].

**Table 6**. Rate parameters for the unimolecular initiation of cyclobutane at P=1 atm, 600 ≤ T (K) ≤ 2000 K and related to Scheme 5.

|  | $k_{1-4}$ | $k_{4-1}$ | $k_{4-5}$ | $k_{4-6}$ | $k_{4-7}$ | $k_{4-8}$ |
| --- | --- | --- | --- | --- | --- | --- |
| log A (s$^{-1}$) | 18.11 | 9.89 | 6.77 | 0.51 | 9.78 | -3.07 |
| n | -0.466 | 0.311 | 1.480 | 3.015 | 1.1 | 4.157 |
| $E_a$ (kcal/mol) | 85.18 | 1.7 | 7.76 | 17.78 | 26.16 | 32.43 |

However, the analysis did not take into account transition state and Tsang used a constant value of 0.16 for the ratio $k_{4-5}/ k_{4-1}$ in accordance with the disproportionation to combination ratio for n-propyl radicals. At 1100 K, the ratios obtained between our values and those proposed by Tsang for $k_{1-4}$, $k_{4-1}$, $k_{4-5}$ and $k_{4-7}$ are, respectively, 0.5, 1.5, 1.6 and 1.7. Despite uncertainties in both experimental and theoretical calculations, the agreement is therefore quite satisfying. It is also possible to compare some computed rate parameters for elementary reactions in Scheme 5 with values obtained using semi-empirical relations as proposed by O'Neal [51]. Accordingly, the activation energy involved in reaction 4 → 6, corresponding to a H-transfer (disproportionation), can be estimated by the following relationship [51]:

$$E_a = E_D + E_{SE} \tag{13}$$

where $E_D$ represents the activation energy involved in the disproportionation of two alkyl free radicals and $E_{SE}$ represents the strain energy of the cyclic transition state. The latter was taken to be 6.3 kcal.mol$^{-1}$ since the transition state exhibits a five-member ring [29]. An activation energy of 1 kcal.mol$^{-1}$ can be reasonably taken into account for disproportionation [49]. From our calculations the activation enthalpy of reaction 4→ 6, is estimated to 6 kcal.mol$^{-1}$ at 298 K and therefore it is in good agreement with the semi-empirical estimation. On the other hand, our quantum calculations give an activation energy equal to 26.6 kcal.mol$^{-1}$ at 298 K for the β-scission of biradical 4 (reaction 4→7). This value can be compared with the mean activation energy for the β-scission of a C-C bond in an alkyl free radical (28.7 kcal.mol$^{-1}$ [49]).

In Scheme 5, biradicals 4 represents the most stable conformer of the biradical although it does not lead directly to the formation of 1-pentene. Since some activation energies, such as that of the reaction 4 → 5 ($\Delta H^{\neq}$ = 6.7 kcal.mol$^{-1}$ at 298 K), have values close to that of the rotation barrier heights, the role of rotational hindrance has been examined in the case of cyclopentane. **Scheme 6** contains the detailed mechanism.

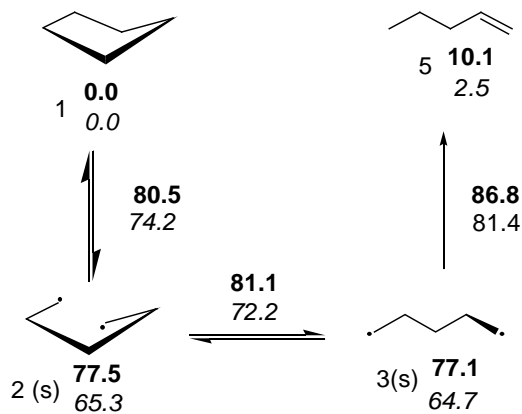

(s) singlet state

$\Delta G°(T)$ in kcal.mol$^{-1}$ (bold : T= 298K, italic : T=1000K)

**Scheme 6**. Detailed mechanism of 1-pentene formation from the ring opening of cyclopentane and obtained by considering different conformers of C$_5$ biradicals.

Two conformations of the biradical •C$_5$H$_{10}$• are involved in Scheme 6 but only one (biradical 3) leads to the formation of 1-pentene by H-abstraction. This result has been validated by IRC calculations. Geometries of TS$_{3-5}$ and biradicals (2) and (3) are presented in **Figure 2**.

As shown, the distance between carbon atom 2 and hydrogen atom 11 is larger in biradical 2 than in biradical 3 (compares structures a and c), which renders H-abstraction for this last species more favorable. This result can be explained by a *gauche* interaction in biradical 2 involving carbon atoms 4 and 1. In biradical 3, carbon atoms 1, 3, 4 and 5 are in the same plan and no *gauche* interaction between carbon atoms 1 and 4 is possible.

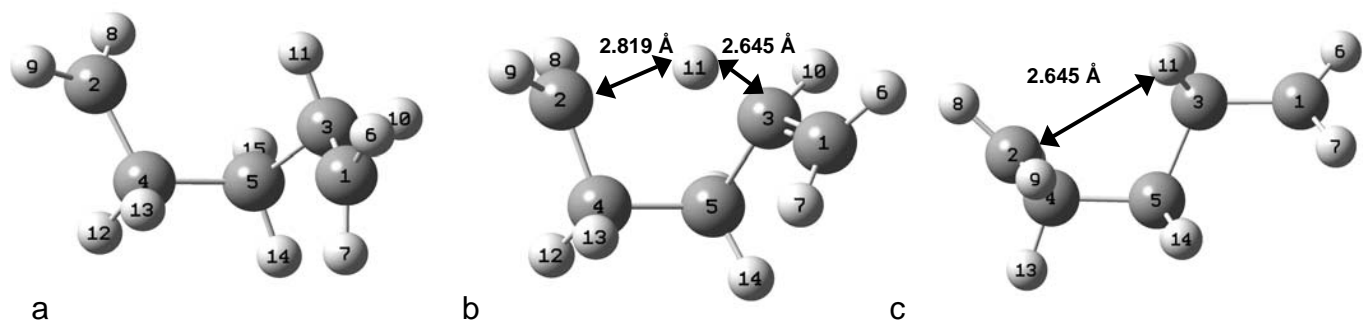

**Figure 2.** Geometries relative to biradical 2 (a), and TS$_{3-5}$ (b), biradical 3 (c)

It is worth noting the different role played by rotational hindrance in cyclopentane and cyclobutane. Indeed, in cyclobutane (Scheme 4), the rotational barrier is of the same order of magnitude as the activation energy for β-scission and both conformers may lead to the formation of ethylene molecules. In cyclopentane, the rotational barrier is lower than the activation energy for 1-pentene formation and only one conformation (biradical 3) allows the latter process. **Table 7** summarizes the rate parameters for the mechanism shown in Scheme 6.

**Table 7.** Rate parameters for the unimolecular initiation of cyclopentane at P=1 atm, 600 ≤ T (K) ≤ 2000 K and related to Scheme 6.

|  | $k_{1-2}$ | $k_{2-1}$ | $k_{2-3}$ | $k_{3-2}$ | $k_{3-5}$ |
|---|---|---|---|---|---|
| log A (s$^{-1}$) | 18.11 | 10.74 | 10.97 | 10.54 | 6.89 |
| n | -0.466 | 0.207 | 0.569 | 0.602 | 1.494 |
| E$_a$ (kcal/mol) | 84.76 | 1.7 | 2.68 | 2.97 | 7.45 |

In order to validate the rate parameters involved in Schemes 5 and 6 and to quantify the effect of the rotational barriers between biradicals 2 and 3, we have compared the global rate constants obtained experimentally by Tsang [25] for the following reactions:

cyclopentane → 1-pentene,  k$_1$= 10$^{16.1}$ exp(-84840 (cal.mol$^{-1}$)/RT),

cyclopentane → cyclopropane + ethylene,  k$_2$= 10$^{16.25}$ exp(-95060 (cal.mol$^{-1}$)/RT),

with global rate constant estimated by QSSA performed on biradical 4 (Scheme 5) and biradicals 2 and 3 (Scheme 6). The rate expressions are:

$$k_{g,1-5}^{scheme\,5} = \frac{k_{1-4}\,k_{4-5}}{k_{4-1} + k_{4-5}}, \tag{14}$$

$$k_{g,1-5}^{scheme\,6} = \frac{k_{1-2}\,k_{2-3}\,k_{3-5}}{k_{2-1}\,k_{3-2} + k_{2-1}\,k_{3-5} + k_{2-3}\,k_{3-5}}, \tag{15}$$

$$k_{g,1-7}^{scheme\,5} = \frac{k_{1-4}\,k_{4-7}}{k_{4-1} + k_{4-7}}, \tag{16}$$

and the kinetics parameters are given in **Table 8**.

**Table 8**. Rate parameters for the global reactions $cC_5H_{10} \rightarrow$ 1-pentene and $cC_5H_{10} \rightarrow cC_3H_6 + C_2H_4$ at P=1 atm, $600 \leq T\,(K) \leq 2000$ K and related to Schemes 5 and 6.

|  | $k_{g(1-5)}^{scheme5}$ | $k_{g(1-7)}^{scheme5}$ | $k_{g(1-5)}^{scheme6}$ |
|---|---|---|---|
| log A (s$^{-1}$) | 20.39 | 24.33 | 20.06 |
| n | -0.970 | -1.542 | -0.878 |
| E$_a$ (kcal/mol) | 92.86 | 112.49 | 92.23 |

Comparison of global rate constants estimated by the QSSA approach and those proposed by Tsang are presented in **Figure 3**. In the range 1000K –1200K, the values calculated from our global rate constants for the formation of 1-pentene ($k_{g(1-5)}^{scheme5}$ and $k_{g(1-5)}^{scheme6}$) are in a good agreement with those obtained by Tsang, our values being lower by a factor 1.2 to 2.

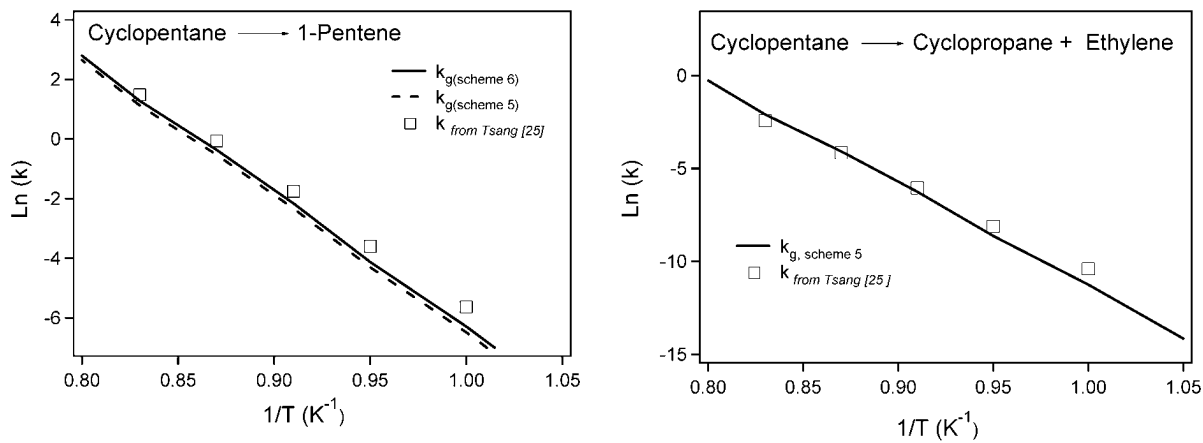

**Figure 3**. Comparison between calculated rate constant and experimental data for the global reactions c-$C_5H_{10} \rightarrow$ 1-pentene and c$C_5H_{10} \rightarrow$ c-$C_3H_6$ + $C_2H_4$

A similar result has been obtained in the case of the reaction leading to cyclopropane and ethylene, with values underestimated by a factor 0.7 to 2.5 with respect to Tsang estimations. Concerning the influence of rotational hindrance in the formation of 1-pentene, $k_{g(1-5)}^{scheme5}$ is always found to be lower than $k_{g(1-5)}^{scheme6}$ over the temperature range (28 % at 600K and 8% at 2000K). By analogy with cyclobutane, calculations performed by considering the different conformations for biradical •$C_5H_{10}$• permit to obtain results closer to experiment. Moreover, the effect of rotational hindrance is slightly greater in the case of cyclopentane than cyclobutane. As said above, this can be explained by the fact that biradical •$C_5H_{10}$• must necessarily rotate in order to yield 1-pentene whereas in cyclobutane, both *gauche* and *trans* conformations of the biradical can decompose in two ethylene molecules.

### 5.3 Cyclohexane

Only a few numbers of studies have been performed on the unimolecular initiation mechanism of cylohexane [26, 28]. **Scheme 7** shows our results for the ring opening of c-$C_6H_{12}$ and subsequent reactions. As before, in this scheme, we only consider the lowest free enthalpy conformers of the biradicals. **Table 9** summarizes the computed rate parameters.

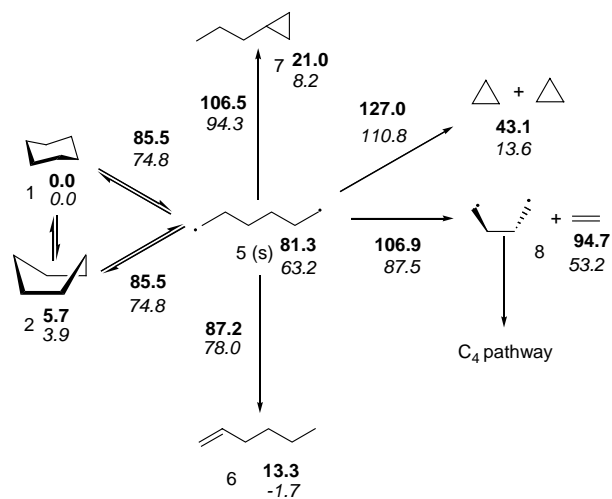

(s) singlet state

$\Delta G°(T)$ in kcal.mol$^{-1}$ (bold : T= 298K, italic : T=1000K)

**Scheme 7.** Mechanism obtained for the ring opening of cyclohexane and for conformers of lowest energies.

**Table 9**. Rate parameters for the unimolecular initiation of cyclohexane at P=1 atm, $600 \leq T (K) \leq 2000$ K and related to scheme 7.

|  | $k_{1-5}$ | $k_{5-1}$ | $k_{2-5}$ | $k_{5-2}$ | $k_{5-6}$ | $k_{5-7}$ | $k_{5-8}$ | $k_{5\text{-c-C3H6}}$ |
|---|---|---|---|---|---|---|---|---|
| log(A s$^{-1}$) | 21.32 | 9.91 | 20.11 | 10.38 | 2.46 | -1.33 | 10.40 | 5.23 |
| n | -0.972 | 0.136 | -0.785 | 0.137 | 2.569 | 3.800 | 0.994 | 2.185 |
| $E_a$ (kcal/mol) | 92.63 | 2.09 | 85.77 | 2.13 | 1.42 | 17.22 | 25.75 | 44.25 |

In our study, the chair and boat conformations of cyclohexane have been taken into account. As mentioned by Dixon and al. [52], the boat structure ($C_{2v}$ symmetry) is a transition state that connects the chair structure with a twist boat ($D_2$ symmetry) conformation. We did not locate the transition state corresponding to the $C_{2v}$ boat structure but we have obtained a $D_2$ twist boat conformation that does correspond to a characterized energy minimum. Hereafter, this twist boat conformation will be simply refereed to as the "boat conformation" (symmetry number of 4, $D_2$ symmetry). At high temperature, the concentration of this boat conformation cannot be neglected and the equilibrium constant $K_{eq}$, corresponding to the reaction:

$$\text{c-C}_6\text{H}_{12} \text{ (chair)} \rightleftharpoons \text{c-C}_6\text{H}_{12} \text{ (boat)}$$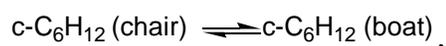,

has been fitted in the 600 - 2000 K temperature range using the relation:

$$K_{eq} = \exp(-\frac{\Delta H_r^\circ}{RT} + \frac{\Delta S_r^\circ}{R}) = \exp(-\frac{3265}{T} + 1.271) \quad (17)$$

CBS-QB3 calculations performed from chair and boat conformations of cyclohexane and considerations of isodesmic reactions have permitted to calculate $\Delta S_r^\circ$ and $\Delta H_r^\circ$. The mean values for $\Delta H_r^\circ$ and $\Delta S_r^\circ$ are, respectively, 6.5 kcal.mol$^{-1}$ and 2.5 cal.mol.l$^{-1}$. From equation (17), K$_{eq}$ is equal to 0.047 at 753K that corresponds to 95.5 % of cyclohexane molecules in chair conformation. This result is in very good agreement with the value reported by Walker and Gulati [53] who reported a slightly larger value (99.5%). Another estimation of the equilibrium constant has been reported by Eliel and Wilen [54] from experimental values obtained at 1073K. They found that 25% of the twist boat structure is present in the mixture, that leads to K$_{eq}$ = 0.33. From equation (17), we obtain K$_{eq}$ = 0.17 at 1073K that is consistent with the estimation made by Eliel and Wilen. Thus, it appears that, in the range of temperature considered in our study, the chair/boat conformation ratio must be taken into account in kinetic calculations.

Two different TSs have been found for the ring opening of cyclohexane depending on its initial boat or chair structure (**Figure 4**), the TS corresponding to the former conformation being 2.4 kcal.mol$^{-1}$ lower in free energy. Since activation energy for ring opening is much higher than the energy involved in boat/chair conformation change, we conclude that only the lowest TS should be taken into account in the kinetic scheme.

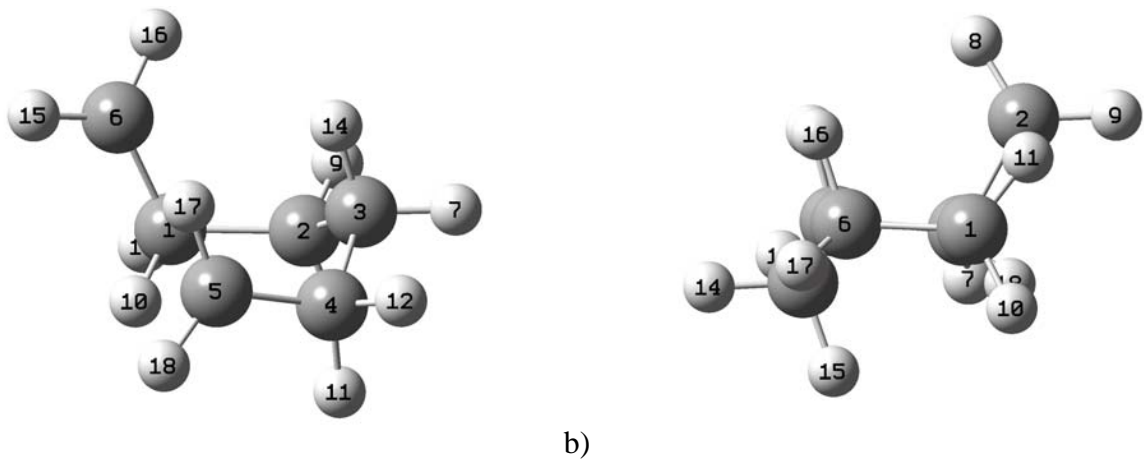

a)  b)

**Figure 4**: TS obtained from the ring opening of cyclohexane with the boat conformation (a) and the chair conformation (b)

The free enthalpy of activation obtained for the ring opening of cyclohexane ($\Delta G^{\neq}$ = 85.5 kcal.mol$^{-1}$, at T = 298K) is higher than that obtained for cyclobutane ($\Delta G^{\neq}$ = 60.7 kcal.mol$^{-1}$) or cyclopentane ($\Delta G^{\neq}$ = 80.5 kcal.mol$^{-1}$) at the same temperature, and close to the value found for a linear alkane. This result will be discussed in detail below but it is interesting to point out that it is consistent with an "unstrained structure" in cyclohexane. The main way of decomposition for the biradical •C$_6$H$_{12}$• (biradical 5) is the formation of 1-hexene, as previously mentioned by Tsang [26]. Indeed, this process involves a slightly constrained transition state for H-abstraction (six-member ring) with a low activation energy $\Delta H^{\neq}$ = 1.8 kcal.mol$^{-1}$ (vs 17.6 kcal.mol$^{-1}$ for c-C$_4$H$_8$ and 6 kcal.mol$^{-1}$ for c-C$_5$H$_{10}$ at 298 K). The value of $\Delta H^{\neq}$ = 1.8 kcal.mol$^{-1}$ can be compared with the semi-empirical estimation of $E_a$ given by equation 13 and corresponding to a disproportionation process. Taking 1 kcal.mol$^{-1}$ for $E_D$ and 0.7 kcal.mol$^{-1}$ for $E_{RS}$ (disproportionation of two alkyl radicals for $E_D$ and a ring strain energy of a six-member ring for $E_{RS}$), $E_a$ = 1.7 kcal.mol$^{-1}$, in very good agreement with our calculation. Owing to this very low activation energy, other routes for •C$_6$H$_{12}$• decomposition are unlikely. An interesting remark can be made for β-scission leading to the formation of •C$_4$H$_8$• and ethylene (reaction 5 → 8). In the experimental study performed by Tsang [26], no cyclobutane was detected what could implicitly be explained by a very low reaction rate for •C$_4$H$_8$• formation compared to 1-hexene. Analyzing free enthalpies of activation at T=1000K in Scheme 7 shows that the ratio between the disproportionation leading to 1-hexene (reaction 5→ 6) and the β-scission (reaction 5→ 8) is about 120. A comparable value is obtained in cylopentane decomposition (Scheme 5). However, in the case of the thermal decomposition of cyclopentane, cylopropane is detected experimentally [25] (though it represents a minor product) whereas for cylohexane, no cyclobutane is observed at all. This discrepancy can be explained by the very fast decomposition of the biradical •C$_4$H$_8$• in two ethylene molecules, compared to the cyclization reaction, as shown in Scheme 3. This assumption cannot be verified using the experimental data reported by Tsang [26] since ethylene formed by initiation would represent a minor part of the total concentration of this molecule, which is principally obtained in propagation reactions (decomposition of 1-hexene and cylohexyl radical). Activation energy of the β-scission of biradical 5 (reaction 5→ 8) is equal to 25.2 kcal.mol$^{-1}$ at 298 K, which is consistent with the β-scission of a C-C bond in an alkyl free radical, with a corresponding activation energy of 28.7 kcal.mol$^{-1}$ [50].

Let us now consider the effect of conformers of •C$_6$H$_{12}$•, that was ignored in Scheme 7. The activation energy involved in the formation of 1-hexene from the biradical •C$_6$H$_{12}$• is quite low ($\Delta H^{\neq} =$ 1 kcal.mol$^{-1}$ at 298 K) and might be competitive with rotational barriers. In accordance with this assumption, we developed a detailed mechanism for the formation of 1-hexene involving the different conformations of the biradical (**Scheme 8**).

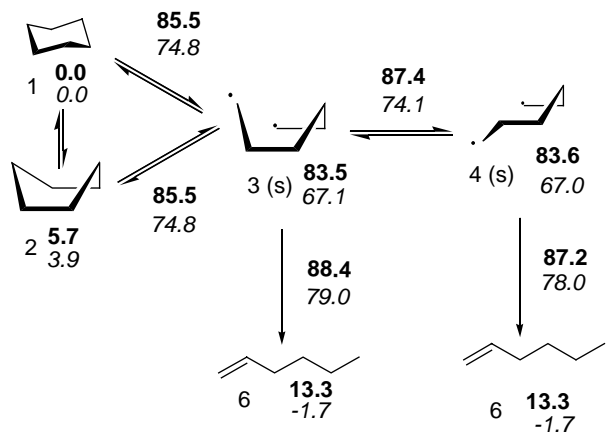

(s) singlet state

$\Delta G°(T)$ in kcal.mol$^{-1}$ (bold : T= 298K, italic : T=1000K)

**Scheme 8**. Detailed mechanism of 1-hexene formation from the ring opening of cyclohexane and obtained by considering different conformers of $C_6$ biradicals.

**Table 10**. Rate parameters for the unimolecular initiation of cyclohexane at P=1 atm, 600 ≤ T (K) ≤ 2000 K and related to scheme 8.

|  | $k_{1-3}$ | $k_{3-1}$ | $k_{2-3}$ | $k_{3-2}$ | $k_{3-6}$ | $k_{3-4}$ | $k_{4-3}$ | $k_{4-6}$ |
|---|---|---|---|---|---|---|---|---|
| log A (s$^{-1}$) | 20.68 | 11.08 | 20.16 | 11.23 | 2.17 | 10.11 | 11.40 | 4.17 |
| n | -0.799 | 0.117 | -0.810 | 0.073 | 2.923 | 0.720 | 0.359 | 2.295 |
| $E_a$ (kcal/mol) | 92.44 | 0.70 | 85.99 | 0.81 | 0.77 | 2.50 | 2.98 | 0.20 |

**Table 10** summarizes the rate parameters for elementary reactions in Scheme 8. In order to compare the global rate constant obtained by Tsang for the thermal decomposition of cyclohexane in 1-hexene [26] to that obtained from Schemes 7 and 8, we performed quasi-stationary-state approximation (QSSA) on biradical 5 (Scheme 7) and biradicals 3 and 4 (Scheme 8). Scheme 8 leads to the following expressions for the reaction c-$C_6H_{12}$ → 1-$C_6H_{12}$ :

$$k_{chair}^{scheme8} = \frac{k_{3-6} k_{1-3}}{C} + \frac{k_{4-6} k_{3-4} k_{1-3}}{C (k_{4-3} + k_{4-6})} \tag{18}$$

$$k_{boat}^{scheme8} = \frac{k_{3-6} k_{2-3}}{C} + \frac{k_{4-6} k_{3-4} k_{2-3}}{C (k_{4-3} + k_{4-6})} \tag{19}$$

$$\text{with } C = \frac{k_{3-4} k_{4-6}}{k_{4-3} + k_{4-6}} + k_{3-1} + k_{3-2} + k_{3-6} \tag{20}$$

At a given temperature, the global rate constant $k_{g(1-6)}^{scheme8}$ can be calculated from rate constants given by equations (18) and (19) and equilibrium constant in order to have the ratio of boat and chair conformations. Thus, we obtained the following expression for the global rate constant:

$$k_{g(1-6)}^{scheme8} = \frac{k_{chair}^{scheme8}}{1+K_{eq}} + \frac{K_{eq}\, k_{boat}^{scheme8}}{1+K_{eq}} \qquad (21)$$

where $K_{eq}$ is the equilibrium constant obtained from equation (17).

For the more general Scheme 7, QSSA applied on the biradical 5 leads to the following relations:

$$k_{chair}^{scheme7} = \frac{k_{1-5}\, k_{5-6}}{k_{5-1}+k_{5-6}} \quad \text{and} \quad k_{boat}^{scheme7} = \frac{k_{2-5}\, k_{5-6}}{k_{5-2}+k_{5-6}} \qquad (22)$$

By analogy with equation (21), the global rate constant for Scheme 7 can be the calculated by using the equilibrium constant:

$$k_{g(1-6)}^{scheme7} = \frac{k_{chair}^{scheme7}}{1+K_{eq}} + \frac{K_{eq}\, k_{boat}^{scheme7}}{1+K_{eq}} \qquad (23)$$

Rate parameters obtained by fitting $k_{g(1-6)}^{scheme7}$ and $k_{g(1-6)}^{scheme8}$ for temperatures ranging from 600K to 2000K are presented in **Table 11**.

**Table 11**. Rate parameters for the global reactions $cC_6H_{12} \to$ 1-hexene at P=1 atm, $600 \leq T\,(K) \leq 2000$ K and related to Schemes 7 and 8.

|  | $k_{g(1-6)}^{scheme7}$ | $k_{g(1-6)}^{scheme8}$ |
|---|---|---|
| log A (s$^{-1}$) | 20.45 | 20.29 |
| n | -0.685 | -0.639 |
| $E_a$ (kcal/mol) | 93.01 | 94.52 |

**Figure 5** shows the comparison of the values calculated from $k_{g(1-6)}^{scheme7}$ and $k_{g(1-6)}^{scheme8}$ (Table 11) and those obtained from the rate constant proposed by Tsang [26], for temperatures between 950K and 1100K, i.e. in the range of validity of Tsang's study.

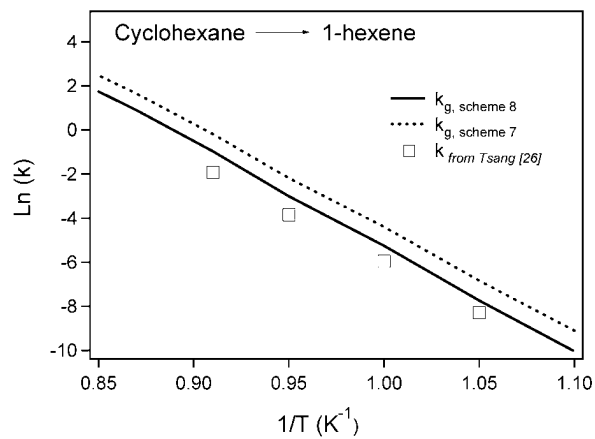

**Figure 5**: comparison between calculated rate constant and experimental data [26] for the global reactions c-$C_6H_{12}$ → 1-hexene.

The agreement between $k_{g(1-6)}^{scheme\,8}$ and the rate constant proposed by Tsang is rather satisfactory since the ratio $\dfrac{k_{g(1-6)}^{scheme\,8}}{k_{Tsang}}$ lies between 2 and 2.6. Another point concerns the difference obtained between the rate constants $k_{g(1-6)}^{scheme\,7}$ and $k_{g(1-6)}^{scheme\,8}$ shown in Figure 6. By neglecting rotational hindrance (Scheme 7), the global rate constant $k_{g(1-6)}^{scheme\,7}$ is overestimated by a factor 2 compared to $k_{g(1-6)}^{scheme\,8}$, in the temperature range 950 – 1100 K. This difference can be explained by the low activation energy involved in the formation of 1-hexene compared to rotational barrier, but also by entropic effects due to the difference between entropy of the "linear" biradical (biradical 5) and biradicals 3 and 4 (Table 1). This results shows that by considering only the lowest energy biradical conformation, the global rate constant is largely overestimated.

### 5.4 Ring strain energies

The results obtained for the ring opening of cycloalkanes into biradicals, show an increase of the enthalpy of activation going from cyclobutane to cyclohexane. These differences are mainly due to the change of ring strain energy when one goes from cycloalkanes to TS. To discuss the results obtained, we first calculated the ring strain energy of cycloalkanes by using the usual definition of RSE [55] (equation 24) :

$$\text{RSE} = \Delta H_{RS} = H_{cyclo} - n\, H_{CH2} \qquad (24)$$

where $H_{cyclo}$ and $H_{CH2}$, represent, respectively, the electronic enthalpies of cycloalkane and of $CH_2$ fragment in a strain-free alkane. *n* represents the number of $CH_2$ fragments in the cyclic species. $H_{CH2}$ has been calculated by difference between the enthalpy of *n*-hexane and *n*-pentane. $H_{cyclo}$ and $H_{CH2}$ have been obtained by considering the electronic energy, zero-point energy and thermal corrections to enthalpy given at a CBS-QB3 level of theory at 298K.

RSE can also be obtained by using enthalpies calculated by isodesmic reactions. Thus, equation 24 can be rewritten as :

$$\text{RSE} = \Delta H_{RS} = \Delta_f H°_{cyclo} - n\, \Delta_f H°_{CH2} \qquad (25)$$

where $\Delta_f H°_{cyclo}$ and $\Delta_f H°_{CH2}$ correspond, respectively, to the enthalpy of cycloalkane formation obtained from isodesmic reactions and enthalpy of a $CH_2$ group formation, obtained by the difference between the enthalpy of formation of *n*-hexane and *n*-pentane.

The values obtained for cyclobutane, cyclopentane, cyclohexane (twist boat) and cyclohexane (chair) are summarized in **Table 12**.

RSE obtained at a CBS-QB3 level from direct electronic energies and those obtained from isodesmic reactions are in a very good agreement with values proposed by Cohen [56] and based on group additivity method.

**Table 12**. Ring strain energies (in kcal.mol$^{-1}$) of cycloalkanes as calculated at the CBS-QB3 level of theory and those proposed by Cohen [56] at T=298K.

| Cycloalkane | Cyclobutane | Cyclopentane | Cyclohexane (Chair) | Cyclohexane (Twist Boat) |
|---|---|---|---|---|
| RSE given by Electronic energies | 27.0 | 7.4 | 1.1 | 7.5 |
| RSE calculated from Isodesmic reactions | 26.8 | 7.5 | 1.0 | 7.5 |
| RSE from Cohen [56] | 26.8 | 7.1 | 0.7 | / |

It is now interesting to estimate the remaining part of RSE contained in the transition states during ring opening. This can be done by comparing the enthalpies of activation obtained for the ring opening of cycloalkanes with those obtained for the dissociation of free-strain linear alkanes and by assuming that the difference obtained between the cyclic species and the corresponding linear one is only due to the ring strain energy. If no difference is observed, one can conclude that no remaining RSE is contained in the TS.

For a linear alkane, the bond dissociation energy (BDE) can be assimilated with the activation enthalpy since the recombination of the two free radicals is barrierless. BDE between two secondary carbon atoms have been estimated by CBS-QB3 methods for *n*-butane, *n*-pentane and *n*-hexane by means of reaction (26):

$$C_nH_{2n+2} \rightarrow x \; ^\bullet C_3H_7 + y \; ^\bullet C_2H_5 \qquad (26)$$

where $C_nH_{2n+2}$ represents a linear alkane; $x$ and $y$ depend on the value of $n$.

BDE corresponds to the enthalpy of reaction (24) and can be expressed by the following equation:

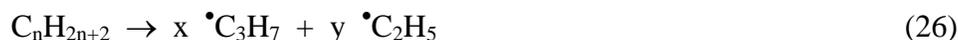

$$BDE = \Delta_r H° = x \; H_{C3H7} + y \; H_{C2H5} - H_{CnH2n+2} \approx \Delta H^\neq \qquad (27)$$

where $H_{C3H7}$, $H_{C2H5}$, represent, respectively, electronic enthalpy of n-propyl and ethyl radicals, and $H_{CnH2n+2}$, the enthalpy of n-butane, n-pentane or n-hexane.

Calculations of BDE in equation (27) can be done using electronic enthalpies or enthalpies estimated from isodesmic reaction. **Table 13** gives the results obtained in the two cases and the experimental values proposed by Luo [57].

**Table 13**: BDE (in kcal.mol$^{-1}$) of two secondary carbon atoms for unstrained linear alkanes, at 298 K

| BDE (C-C) | CBSQ-B3 Electronic energies | CBSQ-B3 Isodesmic reactions | Experimentals BDE [57] |
|---|---|---|---|
| n-Butane | 88.8 | 87.0 | 86.8 |
| n-Pentane | 89.6 | 87.9 | 87.3 |
| n-Hexane | 90.5 | 88.9 | 87.5 |

BDE calculated from direct electronic calculations are higher that experimental ones [57]. On the other hand, the values deduced from isodesmic calculations are closer to these recommended values and show the best accuracy obtained with isodesmic reactions. If we consider that BDE can be assimilated to enthalpy of reaction in the case of linear unstrained alkane, we can compare the activation energies obtained for the opening of cycloalkanes and those obtained by removing the RSE from the BDE of the corresponding linear free-strain alkane. **Table 14** shows this comparison from isodesmic calculations.

**Table 14**. Comparison of $\Delta H^{\neq}$ (in kcal.mol$^{-1}$), obtained for the ring opening of cycloalkanes and those estimated from linear unstrained alkane, at T=298K by taking into account isodesmic reactions.

| Species | $\Delta H^{\neq}$ for cycloalkane | Ring strain energy (RSE) | $\Delta H^{\neq}$ calculated from the corresponding linear alkane | Remaining RSE |
|---|---|---|---|---|
| Cyclobutane | 61.7 | 26.8 | 87 – 26.8 = 60.2 | 1.5 |
| Cyclopentane | 82.4 | 7.5 | 87.9 – 7.5 = 80.4 | 2.0 |
| Cyclohexane (chair) | 89.5 | 1.0 | 88.9 – 1.0 = 87.9 | 1.6 |
| Cyclohexane (twist boat) | 83.0 | 7.5 | 88.9 – 7.5 = 81.4 | 1.6 |

The remaining RSE shown in table 14 permits to conclude that almost all of the RSE is removed in the transition state of cyloalkanes, excepted, perhaps, for cyclohexane (chair) where isodesmic calculations show a slightly increase of the ring strain energy in the transition state. **Figure 6** shows the three transition state involved in the ring opening of the cycloalkanes studied.

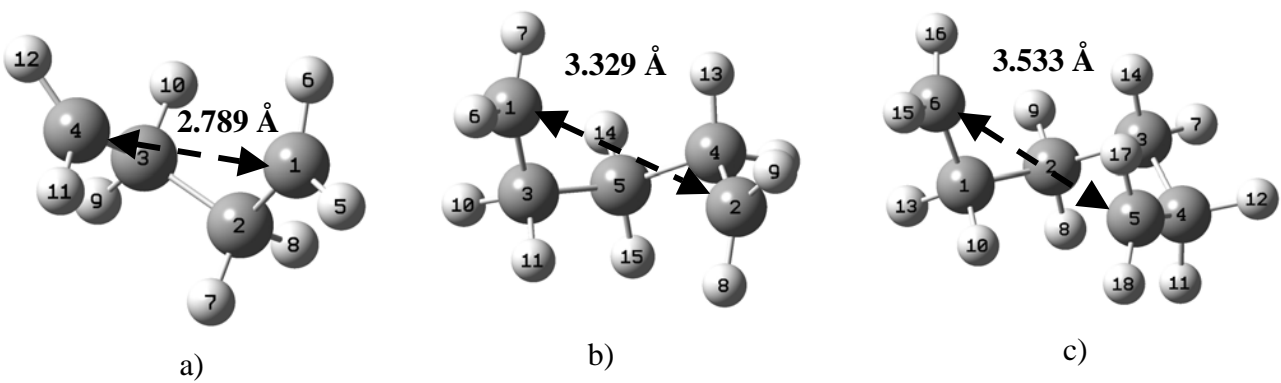

**Figure 6**: TS involved in the ring opening of cyclobutane (a), cyclopentane (b) and cyclohexane (c)

Dudev et al. [55] have showed, from ab initio calculations, that the ring strain energy of cycloalkanes can be explained by contributions of ring bond angles, bond lengths or dihedral angles, this last parameter reflecting nonbonded interactions. Thus, the ring strain energy remaining in the TS of cyclohexane can be explained by the fact that the six-member ring is forced to adopt an energetically unfavorable *gauche* conformation, whereas n-hexane can exist in a strain-free *trans-trans-trans* conformation. Moreover, they showed that the effects of ring bond angles on RSE decrease when the size of the cycle increases while an opposite effect is obtained for nonbonded interactions. In our study, the formation of TS from cyclobutane involves an increase of the ring bond angles which permits to remove a large part of RSE. Indeed, in the TS of Figure 6a, $\angle C_1C_2C_3 = \angle C_2C_3C_4 = 108.4°$ vs $88.6°$ in cyclobutane. For TS involved in the ring opening of cyclopentane (Figure 6b), the large increase of the bond length between $C_1$ and $C_2$ (3.329 Å in TS vs 1.545 Å in cyclopentane) associated with an increase of the ring bond angle ($\angle C_1C_3C_5 = \angle C_2C_4C_5 = 114.8°$ in the TS vs $104.8°$ in the molecule), may explain a large part of the removal of RSE in the TS. In cyclohexane, the low value of RSE is maintained in the transition state, that can be explained by *gauche* interactions remaining in the TS.

**6. Conclusions**

While much work have been carried out on the thermal reactions of aliphatic hydrocarbons, the high temperature reactions of cyclic hydrocarbons has not been explored extensively. In this study, the ring opening of the most representative cyclic alkanes, i.e. cyclobutane, cyclopentane, and cyclohexane have been extensively explored by means of quantum chemistry. All the possible elementary reactions have been investigated from the biradicals yielded by the initiation steps. The thermochemical properties of all the species have been calculated with the high-level CBS-QB3 method. The inharmonic contribution of hindered rotors have been taken into account and isodesmic reactions have been systematically used for the evaluation of the enthalpies in order to minimize the systematic errors. The enthalpies of formation of the biradicals have been compared to data obtained with a semi-empirical method and show a very good agreement.

For all the elementary reactions, the transition state theory allowed to calculate the rate constant. Three parameters Arrhenius expressions have been derived in the temperature range 600 to 2000 K at atmospheric pressure. Tunneling effect has been taken into account in the case of internal H transfers. Thanks to the Quasi Steady State Approximation applied to the biradicals, rate constants have been calculated for the global reaction leading directly from the cyclic alkane to the molecular products. These values have been compared with the few data available in the literature and showed a rather good agreement. The main reaction routes are the decomposition to two ethylene molecules in the case of cyclobutane and the internal disproportionation of the biradicals yielding 1-pentene and 1-hexene in the case of cyclopentene and cyclohexane, respectively. An important fact highlighted in this work is the role of the internal rotation hindrance in the biradical fate. Whereas the energy barriers between conformers are usually of low energy in comparison to the reaction barriers, all the energies are close in this case and taking the rotations between the conformers into account changes the global rate constant especially for the largest biradicals. The analysis of the variation of the ring strain energy has also showed that the larger part is removed when going from the cycloalkanes to the transition states. These last structure are close to be unconstrained with between 1 or 2 kcal remaining.

ACKNOWLEDGMENT

The Centre Informatique National de l'Enseignement Supérieur (CINES) is gratefully acknowledged for allocation of computational resources

SUPPORTING INFORMATION AVAILABLE

The full list of author in ref 30, the structural parameters for all the species investigated in this study the frequencies, energies and zero point energies. This material is available free of charge via the Internet at http://pubs.acs.org.